# Highly Efficient Factorial Designs for cDNA Microarray Experiments: Use of Approximate Theory Together with a Step-up Step-down Procedure

Runchu Zhang
KLAS and School of Mathematics and Statistics
Northeast Normal University
Changchun 130024, China, rczhang@nenu.edu.cn
LPMC and School Of Mathematical Sciences
Nankai University
Tianjin 300071, China, zhrch@nankai.edu.cn

and

Rahul Mukerjee
Indian Institute of Management Calcutta
Joka, Diamond Harbour Road
Kolkata 700 104, India
rmuk0902@gmail.com

**Abstract**: A general method for obtaining highly efficient factorial designs of relatively small sizes is developed for cDNA microarray experiments. The method allows the main effects and interactions of successive orders to be of possibly unequal importance. First, the approximate theory is employed to get an optimal design measure which is then discretized. It is, however, observed that a naïve discretization may fail to yield an exact design of the stipulated size and, even when it yields such an exact design, there is often scope for improvement in efficiency. To address these issues, we propose a step-up/down procedure which is seen to work very well. The resulting highly efficient designs are found to remain almost free from possible dye-color effects under a suitable dye-color assignment. They are also seen to be quite robust to heteroscedasticity as may be caused by biological variability. We focus on the baseline and all-to-next parametrizations but our method works equally well also for hybrids of the two and other parametrizations.

## 1. Introduction

Factorial designs for cDNA microarray experiments have received significant attention in recent years. The fields of application include the biological, agricultural and pharmaceutical sciences. These designs arise when the cell populations under study have a factorial structure, which is often the case. An early example, due to Churchill (2002), concerns a $3 \times 2$ factorial experiment for comparing gene expression in liver tissues of mice from a gallstone-susceptible strain (Pera) and gallstone-resistant strains (DBA and I) on low-fat and high-fat diets. Here the first factor, nature of strain, has three levels Pera, DBA and I, while the second factor, diet, has two levels low-fat and high-fat. Glonek and Solomon (2004) cited another example where the experiment compares two cell lines FIΔ and V449E at times 0 hours and 24 hours. This corresponds to a $2^2$ factorial, where the first factor is the mutant having two levels FIΔ and V449E, and the second factor is time also having two levels 0 and 24 hours.

The objects of interest in a factorial design are the factorial main effects and interactions, as defined via a suitable parametrization. The main effects and interactions of successive orders may not, however, be of equal importance to the experimenter. Taking this possibility into account, we propose a general method for obtaining highly efficient factorial designs of relatively small sizes for cDNA microarray experiments. While the classical literature centers around an orthogonal parametrization (see e.g., Mukerjee and Wu, 2006), we focus on the baseline and all-to-next parametrizations. The baseline parametrization, popular in microarray experiments, is relevant when there is a control or baseline level for each factor. On the other hand, the all-to-next parametrization is attractive when there is a natural ordering of levels for each factor and interest lies in comparing the consecutive levels. More details follow in Section 2. The method proposed here, of course, works well for other parametrizations too, including hybrids of the two that we focus on.

We envisage a very general setup where the numbers of factor levels, the weights specifying the relative importance of the main effects and interactions, and the number of slides available for the experiment can be quite arbitrary. All these make the derivation of exact optimal designs intractable. On the other hand, applied research can hardly wait till the development of a perfect mathematical theory and, in most applications, it suffices to have designs with assured high efficiency, say over 0.95 or 0.90, under the chosen criterion. This is precisely what our general construction method, developed in Section 3, aims at achieving. The method begins by employing the approximate theory, but it is seen that a naïve discretization of the resulting optimal design measure often fails to serve our purpose. In order to overcome this difficulty, we propose a step-up/down procedure which is seen to work very well. Examples are given in Section 4 to demonstrate this.

The possible presence of dye-color effects can be an issue in microarray experiments. Although we begin by ignoring it, it is observed in Section 5 that under a suitable dye-color assignment, our highly efficient designs remain almost free from dye-color effects. The paper ends in Section 6 with some concluding remarks where, among other things, it is seen that our findings remain quite robust to heteroscedasticity as may be caused by biological variability.

In the light of the aforesaid aspects of our work, we now present a literature review and indicate how our approach compares with the existing ones. See Banerjee and Mukerjee (2008), Sanchez (2010) and Schiffl (2011) for more detailed reviews and further references. Considering first the research done under orthogonal parametrization, we refer to Kerr (2006) who investigated economic two-level factorial designs which estimate all main effects and two-factor interactions, without assuming the absence of higher order interactions. Further results on factorial microarray designs under the orthogonal parametrization include those due to Landgrebe et al. (2006), Gupta (2006) and Grossmann and Schwabe (2007). Bueno Filho et al. (2006) also considered a similar parametrization for their model with fixed treatment effects.

More relevant to the present context is the existing work on microarray designs under baseline parameterization which forms a major foundation of this paper. Yang and Speed (2002) introduced this parametrization for $2^2$ factorial designs, while Glonek and Solomon (2004) reported illuminating admissibility results through complete enumeration, primarily for the $2^2$ factorial and also touching upon the $2\times 3$ factorial. The idea of admissibility was followed up by Sanchez and Glonek (2009) and Sanchez (2010), who termed it Pareto optimality and studied Pareto optimal designs for linear functions of the main effects and interactions in $2^2$ and $2\times 3$ factorials under the baseline parametrization, by complete enumeration for smaller designs, and simulated annealing for larger designs. Banerjee and Mukerjee (2008) derived theoretical results on optimal factorial designs under the baseline parametrization. They employed the approximate theory to find weighted optimal designs for $2^2$ factorials but were unable to extend this theory to general factorials. For the latter, they obtained exact optimal designs in the saturated case where no degrees of freedom are left for the error, and proposed certain heuristics for non-saturated cases.

In contrast, as indicated earlier, the main contribution of this paper is the development of a broad-spectrum method, combining the approximate theory with a step-up/down procedure, which yields economic and highly efficient designs for general factorials. These designs are of relatively small sizes but not saturated and hence allow error degrees of freedom for performing tests. Thus our approach caters to a wide variety of situations where mathematical derivation of exact optimal designs is intractable and complete enumeration is infeasible. Moreover, approximate theory provides a benchmark which, though non-attainable in an exact setup, yields an assured lower bound to the efficiency of our designs. This kind of benchmark is not available in simulated annealing.

In designing factorial microarray experiments, approximate theory was used also by Bueno Filho et al. (2006), Grossmann and Schwabe (2007), Passos et al. (2009) and Schiffl (2011), among others. But their settings and criteria and hence final results are different from ours. For example, Bueno Filho et al. (2006) and Grossman and Schwabe (2007) worked under the orthogonal or similar parametrizations, while Passos et al. (2009) applied the approximate theory only to the $3^2$ factorial in a context different from ours. Schiffl (2011) considered parametrizations similar but not identical to the ones studied here. Her optimality results are also different from ours since she explored optimality separately for the main effects and interactions whereas we work with a weighted criterion that includes these factorial effects simultaneously. Finally, none of these authors considered the step-up/down procedure which forms a major component of our approach and, as seen later, leads to a significant gains over a naïve application of the approximate theory.

Before concluding the introduction, we follow Banerjee and Mukerjee (2008) to briefly indicate the experimental setup considered here. In cDNA microarrays, each slide compares two cell populations on the basis of mRNA samples separately labeled with fluorescent dyes, usually red and green. After competitive hybridization, the ratio of the red and green fluorescence intensities is measured at each spot on each slide. Any such ratio represents the relative abundance of the gene in the two cell populations compared on the corresponding slide. The intensity ratios are usually adjusted for background noise and then normalized with the objective of removing systematic biases. We consider linear models for the log intensities and hence the log intensity ratios. In what follows, the modeling and also the corresponding design problem refer to a single gene – it is intended that the same design applies simultaneously to all genes on the array.

**2. Parametrizations, linear model and design criterion**

2.1 *Baseline and all-to-next parametrizations*

To motivate the ideas, consider first the case where a single factor, with levels $0, 1,...,m-1$, dictates the cell populations. Let $\tau(0), \tau(1),...,\tau(m-1)$ denote the expected log intensities, i.e., the effects, of the $m$ levels. The baseline parametrization is relevant when one of the levels, say 0, is the control or baseline level and interest lies in comparing the remaining levels with it. Thus $\theta(0) = \tau(0)$ is the baseline effect and the contrasts of interest are $\theta(j) = \tau(j) - \tau(0)$, $1 \leq j \leq m-1$. Clearly, then

$$\tau(0) = \theta(0) \qquad \text{and} \qquad \tau(j) = \theta(0) + \theta(j), \quad 1 \leq j \leq m-1. \qquad (2.1)$$

The all-to-next parametrization, on the other hand, arises if there is a natural ordering among the levels $0, 1,...,m-1$, as happens, e.g., when the levels denote time points in an increasing order. Interest lies in comparing the consecutive levels. In this case too, we can write $\theta(0) = \tau(0)$ and the contrasts of interest are now $\theta(j) = \tau(j) - \tau(j-1)$, $1 \leq j \leq m$. Thus, instead of (2.1), we now get

$$\tau(j) = \theta(0) + \theta(1) + ... + \theta(j), \qquad 0 \leq j \leq m-1. \qquad (2.2)$$

It is straightforward to extend the baseline and all-to-next parametrizations in (2.1) and (2.2) to the multifactor case. Consider an $m_1 \times ... \times m_n$ factorial involving $n$ factors which dictate the cell populations. Let the levels of the $i$th factor be $0, 1,...,m_i - 1$. Then there are $v = \Pi m_i$ treatment combinations $j_1...j_n$ ($0 \leq j_i \leq m_i - 1$, $1 \leq i \leq n$), each representing a cell population. Let $\tau(j_1...j_n)$ denote the expected log intensity, i.e., the effect, of the treatment combination $j_1...j_n$.

First suppose there is a control or baseline level, say 0, for each factor. Then generalizing (2.1), the baseline parametrization in the multifactor case is given by

$$\tau(j_1...j_n) = \Sigma_{u_1 \in \{0, j_1\}} ... \Sigma_{u_n \in \{0, j_n\}} \theta(u_1...u_n), \qquad (2.3)$$

for each $j_1...j_n$, where $\theta(0...0)$ is the baseline effect and $\theta(u_1...u_n)$ is a main or interaction effect parameter for every $u_1...u_n \neq 0...0$, depending on whether one or more of $u_1,...,u_n$ are nonzero. In (2.3), the sum on each $u_i$ is over $u_i \in \{0, j_i\}$, i.e., $u_i$ is fixed at 0 if $j_i = 0$, while $u_i$ ranges over only 0 and $j_i$ if $j_i > 0$. Thus, for $n = 1$, it is obvious that (2.3) reduces to (2.1).

**Example 1A**. For a $2 \times 3$ factorial, the baseline parametrization in (2.3) can be written explicitly as

$$\tau(00) = \theta(00), \quad \tau(01) = \theta(00) + \theta(01), \quad \tau(02) = \theta(00) + \theta(02), \quad \tau(10) = \theta(00) + \theta(10),$$
$$\tau(11) = \theta(00) + \theta(01) + \theta(10) + \theta(11), \quad \tau(12) = \theta(00) + \theta(02) + \theta(10) + \theta(12).$$

This is precisely how this parametrization is defined in the literature – see e.g., Figure 6 in Glonek and Solomon (2004). Inverting the above equations, here $\theta(10) = \tau(10) - \tau(00)$ represents the main effect of the first factor, while $\theta(01) = \tau(01) - \tau(00)$ and $\theta(02) = \tau(02) - \tau(00)$ represent that of the second factor. Also, the two-factor interaction is represented by $\theta(11) = \tau(11) - \tau(01) - \tau(10) + \tau(00)$ and $\theta(12) = \tau(12) - \tau(02) - \tau(10) + \tau(00)$. □

Continuing with an $m_1 \times ... \times m_n$ factorial, now suppose there is a natural ordering of the levels $0, 1, ..., m_i - 1$ of each factor. Generalizing (2.2), then the all-to-next parametrization in the multifactor case is given by

$$\tau(j_1...j_n) = \Sigma_{u_1=0}^{j_1} ... \Sigma_{u_n=0}^{j_n} \theta(u_1...u_n), \tag{2.4}$$

for each $j_1...j_n$, where $\theta(0...0)$ is analogous to the baseline effect in (2.3) and $\theta(u_1...u_n)$ is a main or interaction effect parameter for every $u_1...u_n \neq 0...0$, according as whether one or more of $u_1,...,u_n$ are nonzero. Note that the sum on each $u_i$ is over $u_i = 0,1,...,j_i$ in (2.4) whereas it is over $u_i \in \{0, j_i\}$ in (2.3). Thus for the $2^n$ factorial, where each $j_i$ is 0 or 1, the two parametrizations become identical, but they are different for other factorials. Clearly, (2.4) reduces to (2.2) if $n = 1$.

**Example 1B**. For a $2 \times 3$ factorial, the all-to-next parametrization in (2.4) becomes

$$\tau(00) = \theta(00), \quad \tau(01) = \theta(00) + \theta(01), \quad \tau(02) = \theta(00) + \theta(01) + \theta(02), \quad \tau(10) = \theta(00) + \theta(10),$$
$$\tau(11) = \theta(00) + \theta(01) + \theta(10) + \theta(11), \quad \tau(12) = \theta(00) + \theta(01) + \theta(02) + \theta(10) + \theta(11) + \theta(12).$$

Inverting these equations, $\theta(10) = \tau(10) - \tau(00)$ represents the main effect of the first factor, while $\theta(01) = \tau(01) - \tau(00)$ and $\theta(02) = \tau(02) - \tau(01)$ represent that of the second factor. Also, the two-factor interaction is represented by $\theta(11) = \tau(11) - \tau(01) - \tau(10) + \tau(00)$ and $\theta(12) = \tau(12) - \tau(02) - \tau(11) + \tau(01)$. Note that now any main effect or interaction parameter reflects a comparison between consecutive levels of the relevant factor(s), whereas in Example 1A, it is relative to the baseline level(s) of the relevant factor(s). □

As Examples 1A and 1B show, the main effect or interaction parameters are contrasts among the treatment effects under both baseline and all-to-next parametrizations. However, these contrasts are not mutually orthogonal. It is here that the two parametrizations differ from the orthogonal parametrization. One can as well think of hybrids of the baseline and all-to-next parametrizations. These are of interest when there is a natural baseline level for some of the factors and a natural ordering of the levels for the other factors. If factors of these two types are called baseline and all-to-next factors respectively, then combining (2.3) and (2.4), such a hybrid parametrization is given by

$$\tau(j_1...j_n) = \Sigma^*_{u_1}...\Sigma^*_{u_n} \theta(u_1...u_n), \qquad (2.5)$$

for each $j_1...j_n$, where for every $i$, the sum $\Sigma^*_{u_i}$ is over $u_i \in \{0, j_i\}$ if the $i$th factor is of the baseline type, or over $u_i = 0,1,...,j_i$ if the $i$th factor is of the all-to-next type. Since $0 \leq j_i \leq m_i - 1$, $1 \leq i \leq n$, it is easy to see that each of (2.3)-(2.5) expresses the effects of the $v = \Pi m_i$ treatment combinations as linear functions of $v = \Pi m_i$ parameters $\theta(u_1...u_n)$. The forms of these linear functions depend on the specific parametrization.

2.2 *Linear model*

In this and the next subsections and also in Section 3, the three parameterizations in (2.3)-(2.5) are covered in a unified manner. We begin by ignoring any dye-color effects – this issue will be taken up in Section 5. Then, for a slide which compares a pair of distinct treatment combinations $(j_1...j_n, j'_1...j'_n)$ in a microarray experiment, the expected log intensity ratio equals $\tau(j_1...j_n) - \tau(j'_1...j'_n)$. Now, under any of the parametrizations (2.3)-(2.5), $\theta(0...0)$ occurs in the expression for every $\tau(j_1...j_n)$. Hence the difference $\tau(j_1...j_n) - \tau(j'_1...j'_n)$ is a linear function of only the $\theta(u_1...u_n)$ with $u_1...u_n \neq 0...0$, i.e., of only the main effect and interaction parameters. Thus if we write $\theta$ for the $(v-1) \times 1$ vector of these parameters, then

$$\tau(j_1...j_n) - \tau(j'_1...j'_n) = x(j_1...j_n; j'_1...j'_n)^T \theta, \qquad (2.6)$$

where the superscript $T$ stands for transposition, and $x(j_1...j_n; j'_1...j'_n)$ is a known $(v-1) \times 1$ vector which depends on $j_1...j_n$, $j'_1...j'_n$ and the specific parametrization.

**Example 1C**. In a $2 \times 3$ factorial, writing $\theta = (\theta(01), \theta(02), \theta(10), \theta(11), \theta(12))^T$, from Example 1A, $\tau(11) - \tau(02) = \theta(01) - \theta(02) + \theta(10) + \theta(11) = x(11;02)^T \theta$, under the baseline parametrization, where $x(11;02) = (1,-1,1,1,0)^T$. Similarly, from Example 1B, $\tau(11) - \tau(02) = -\theta(02) + \theta(10) + \theta(11) = x(11;02)^T \theta$, under the all-to-next parametrization, where $x(11;02) = (0,-1,1,1,0)^T$. □

The $v$ treatment combinations can be paired in $p = \binom{v}{2}$ ways. For notational simplicity, label these $p$ pairs as $1,...,p$ in any fixed order. For any $k$ $(1 \leq k \leq p)$, the log intensity ratio arising from a slide that compares the pair labeled $k$ has expectation $x_k^T \theta$, where by (2.6), $x_k = x(j_1...j_n; j'_1...j'_n)$ if this pair equals $(j_1...j_n, j'_1...j'_n)$. Thus, as illustrated in Example 1C, one can explicitly find the vectors $x_1,...,x_p$ corresponding to the $p$ pairs, taking care of the parametrization adopted.

Now suppose the available resources in a cDNA microarray experiment allow the use of $N$ slides. Consider a design which compares the pairs labeled $k(1),...,k(N)$ on the $N$ slides, where $k(1),...,k(N)$ are not necessarily distinct and $1 \leq k(1),...,k(N) \leq p$. If $Y(i)$ is the observed log intensity ratio from the $i$th slide, then as noted above, $E\{Y(i)\} = x_{k(i)}^T \theta$, $1 \leq i \leq N$. Writing $Y = (Y(1),...,Y(N))^T$ for the observational vector and $X = [x_{k(1)} ... x_{k(N)}]^T$ for the $N \times (v-1)$ design matrix, this leads to the linear model

$$E(Y) = X\theta, \qquad \text{Disp}(Y) = \sigma^2 I_N, \qquad (2.7)$$

where $I_N$ is the identity matrix of order $N$, and $\sigma^2$ is the constant error variance. Here it is assumed that the log intensity ratios arising from different slides have the same variance and are uncorrelated, the latter assumption being justified when independent biological replications are used throughout the experiment; see Section 6 for more discussion. Observe that the information matrix $X^T X$, which plays a crucial role in what follows, can as well be expressed as

$$X^T X = \Sigma_{k=1}^{p} f_k x_k x_k^T, \qquad (2.8)$$

where $f_k$ is the number of slides which compare the pair labeled $k$ ($1 \leq k \leq p$) in the design under consideration. Clearly, the nonnegative integers $f_1,..., f_p$ add up to $N$, the total number of slides.

*2.3 Design criterion*

In microarray experiments not only the main effects but also the interactions are of interest. In fact, the latter are often of greater interest than the former. From this perspective, we consider designs which keep all the main and interaction effect parameters estimable. Let $Q_1$ be the set of main effect parameters and, for $2 \leq i \leq n$, let $Q_i$ be the set of parameters representing the $i$-factor interactions; e.g., in a $2 \times 3$ factorial, $Q_1 = \{\theta(01), \theta(02), \theta(10)\}$ and $Q_2 = \{\theta(11), \theta(12)\}$. With a view to quantifying the relative importance of the main effects and interactions of successive orders, we work with a system of weights $w_1,..., w_n$ such that a weight $w_i$ is attached with each parameter in $Q_i$, $1 \leq i \leq n$. These weights are specified by the experimenter depending on the priorities in a given context. Then our weighted design criterion calls for the minimization of

$$\psi = w_1 S_1 + ... + w_n S_n, \qquad (2.9)$$

where $S_i$ is the sum of variances of the best linear unbiased estimators (BLUEs) of the parameters in $Q_i$, $1 \leq i \leq n$. Thus in a $2 \times 3$ factorial, denoting the BLUE of $\theta(u_1 u_2)$ by $\hat{\theta}(u_1 u_2)$, we have $S_1 = \text{var}\{\hat{\theta}(01)\} + \text{var}\{\hat{\theta}(02)\} + \text{var}\{\hat{\theta}(10)\}$, $S_2 = \text{var}\{\hat{\theta}(11)\} + \text{var}\{\hat{\theta}(12)\}$, and if, for instance, the two-factor interaction is considered twice as important as the main effects then one would take $w_1 = 1$, $w_2 = 2$ in (2.9). Note that if, in particular, $w_1 = ... = w_n$ then $\psi$ reduces to the usual A-criterion.

A matrix formulation for $\psi$ will be helpful. Let $W$ be a diagonal matrix of order $v-1$, such that the $j$th diagonal element of $W$ equals $w_i$ if the $j$th element of $\theta$ belongs to $Q_i$; e.g., in a $2 \times 3$ factorial, with $\theta = (\theta(01), \theta(02), \theta(10), \theta(11), \theta(12))^T$, we have $W = \text{diag}(w_1, w_1, w_1, w_2, w_2)$. Now, if all the main and interaction effect parameters are estimable then by (2.7), $X^T X$ is nonsingular and the dispersion matrix of the BLUE of $\theta$ equals $\sigma^2 (X^T X)^{-1}$. Hence from (2.8) and (2.9), it is not hard to see that

$$\psi = \sigma^2 \text{tr}\{(X^T X)^{-1} W\} = \sigma^2 \text{tr}\{(\Sigma_{k=1}^p f_k x_k x_k^T)^{-1} W\}. \tag{2.10}$$

In view of (2.10), one needs to find the nonnegative integers $f_1, \ldots, f_p$, subject to $f_1 + \ldots + f_p = N$, so as to minimize $\psi$. The discreteness of $f_1, \ldots, f_p$ and the associated combinatorial complexities, however, make this exact design problem intractable, more so because the number of slides $N$, as well as the numbers of levels $m_1, \ldots, m_n$ of the factors and the weights $w_1, \ldots, w_n$ can be quite arbitrary. On the other hand, as we will see in the next section, significant progress can be made via the use of the approximate theory in conjunction with a step-up/down procedure.

## 3. Constructing highly efficient designs

*3.1 Approximate theory*

For $1 \leq k \leq p$, let $\pi_k = f_k / N$ be the proportion of slides which compare the pair labeled $k$, Then $\Sigma_{k=1}^p f_k x_k x_k^T = NM$, where $M = \Sigma_{k=1}^p \pi_k x_k x_k^T$. Hence by (2.10),

$$\psi = (\sigma^2 / N) \text{tr}(M^{-1} W). \tag{3.1}$$

The discreteness of $f_1, \ldots, f_p$ induces the same on $\pi_1, \ldots, \pi_p$, but considerable simplicity is achieved if we invoke the approximate theory and treat $\pi_1, \ldots, \pi_p$ for the time being as nonnegative continuous variables satisfying $\pi_1 + \ldots + \pi_p = 1$. Any such $\pi = (\pi_1, \ldots, \pi_p)$ is called a design measure which assigns masses $\pi_1, \ldots, \pi_p$ on the pairs labeled $1, \ldots, p$, respectively. We write $M = M(\pi)$ to highlight the dependence of $M$ on $\pi$. By (3.1), then the design problem reduces to finding an optimal design measure which minimizes $\text{tr}[\{M(\pi)\}^{-1} W]$ over all possible $\pi$.

A multiplicative algorithm can be applied conveniently to obtain the optimal design measure. It starts with the uniform measure $\pi^{(0)} = (1/p, \ldots, 1/p)$ and finds $\pi^{(h)} = (\pi_1^{(h)}, \ldots, \pi_p^{(h)})$, $h = 1, 2, \ldots$ recursively as

$$\pi_k^{(h)} = \pi_k^{(h-1)} \frac{x_k^T \{M(\pi^{(h-1)})\}^{-1} W \{M(\pi^{(h-1)})\}^{-1} x_k}{\text{tr}[\{M(\pi^{(h-1)})\}^{-1} W]}, \quad 1 \leq k \leq p, \tag{3.2}$$

till a design measure $\pi^{(h)}$, satisfying

$$x_k^T \{M(\pi^{(h)})\}^{-1} W \{M(\pi^{(h)})\}^{-1} x_k - \text{tr}[\{M(\pi^{(h)})\}^{-1} W] \leq \varepsilon, \ 1 \leq k \leq p \qquad (3.3)$$

is reached, where $\varepsilon$ is a preassigned small positive quantity. Let $\tilde{\pi} = (\tilde{\pi}_1, ..., \tilde{\pi}_p)$ denote the terminal design measure $\pi^{(h)}$ which meets (3.3). Then from directional derivative considerations (cf. Silvey, 1980, pp. 19, 35), it can be shown that $\text{tr}[\{M(\tilde{\pi})\}^{-1} W] \leq t_0 + \varepsilon$, where $t_0$ is the minimum possible value of $\text{tr}[\{M(\pi)\}^{-1} W]$ over all possible $\pi$. We take $\varepsilon = 10^{-11}$. Then $\tilde{\pi}$ represents the optimal design measure with computational accuracy as high as up to ten places of decimals. We remark that even at this level of accuracy the algorithm in (3.2) and (3.3), implemented on MATLAB, works very fact under the parametrizations introduced earlier. In all the examples of the next section and many others not reported here, it was seen to terminate almost instantaneously.

*3.2 Improving upon naïve discretization*

An exact design, which has $N$ slides and compares the pairs labeled $1, ..., p$ on $f_1, ..., f_p$ slides respectively, corresponds to the design measure $\pi^{\text{exact}} = (f_1/N, ..., f_p/N)$, and hence its efficiency under our design criterion can be defined as

$$\text{Eff} = \frac{\text{tr}[\{M(\tilde{\pi})\}^{-1} W]}{\text{tr}[\{M(\pi^{\text{exact}})\}^{-1} W]} \qquad (3.4)$$

where $\tilde{\pi}$ is the optimal design measure. Having found $\tilde{\pi} = (\tilde{\pi}_1, ..., \tilde{\pi}_p)$ as in the last subsection, the usual approach for obtaining an efficient exact design from $\tilde{\pi}$ involves multiplying the masses $\tilde{\pi}_1, ..., \tilde{\pi}_p$ by a suitable positive constant, say $c$, and rounding off $c\tilde{\pi}_1, ..., c\tilde{\pi}_p$ to nearest integers, say $f_1, ..., f_p$, respectively. Since $f_1, ..., f_p$ are then approximately proportional to $\tilde{\pi}_1, ..., \tilde{\pi}_p$, one is inclined to believe that an exact design, which compares the pairs labeled $1, ..., p$ on $f_1, ..., f_p$ slides respectively, will be highly efficient. There are, however, two serious difficulties with such naïve discretization of the optimal design measure. These are discussed in (a) and (b) below where we also indicate how a step-up/down procedure, to be introduced formally in the next subsection, helps in resolving these difficulties. Hereafter, for notational simplicity, we write $w = (w_1, ..., w_n)$ to denote the vector of weights attached to the main effects and interactions of successive orders.

(a) First, the aforesaid technique of rounding off to nearest integers may not yield an exact design with a pre-assigned number, $N$, of slides, because of the gaps inherent in discretization. For illustration, consider a $3 \times 5$ factorial under the baseline parametrization. Let $w = (1, 2)$. Then, starting from the optimal design measure $\tilde{\pi}$ and with $c = 35.1305$ and $35.1306$, rounding off yields exact designs, say $d(26)$ and $d(34)$ with $N = 26$ and $34$ slides respectively. Clearly, there is a gap between 26 and 34, i.e., rounding off does not work for $N = 27, ..., 33$. What should one do, for in-stance, with $N = 28$? While arbitrary addition of some slides to $d(26)$ or deletion of some slides from $d(34)$ can have unpredictable consequences, our step-up/down procedure offers a systematic rule with statistical justification. In its simplest version, the procedure finds two exact designs with $N = 28$ slides: one stepping up from $d(26)$ via the addition of one slide to it in each step, and the other stepping down from $d(34)$ via the deletion of one of its slides in each step, the addition or deletion in each step being done in the most efficient manner under the present design criterion. Eventually, in 2 (= 28 – 26) and 6 (= 34 – 28) steps, these step-up and step-down operations yield two 28-slide exact designs, which are seen to have efficiencies 0.9335 and 0.9465 respectively.

(b) Second, even when rounding off to nearest integers yields an exact design with the required number of slides, i.e., the gaps indicated in (a) pose no problem, it may turn out that such an exact design has a singular information matrix $X^T X$, or there may be scope for improving upon its efficiency. This may look somewhat surprising but it is again an feature of discretization. For example, suppose it is required to obtain a 28-slide design for a $2^4$ factorial. Then the baseline and all-to-next parametrizations are equivalent as each factor has two levels. With $w = (1, 2, 2, 1)$, rounding off produces exact designs, say $d(28)$ and $d(48)$, for $N = 28$ and 48. However, the design $d(28)$ is seen to have a singular information matrix. Step-down from $d(48)$ helps and leads to a 28-slide design with efficiency 0.9264. Incidentally, here 48 is the smallest $N$ for which rounding off can produce an exact design having $N$ slides and a nonsingular information matrix. As another example, in a $3^2$ factorial, under the baseline parametrization and with $w = (1, 1)$, rounding off yields exact designs, say $d(18)$, $d(22)$ and $d(30)$, for $N = 18$, 22 and 30. While $d(22)$ has efficiency 0.8974, step-up and step-down from $d(18)$ and $d(30)$ produce two more 22-slide designs which have much higher efficiencies, namely 0.9567 and 0.9608, respectively. Similar examples abound.

*3.3 Step-up step-down procedure*

The *step-up* procedure is employed to obtain an exact design with $N$ slides from an exact design, say $d(N_0)$, which involves $N_0$ ($< N$) slides and has a nonsingular information matrix. It finds, for $i = 0, 1, ..., N - N_0 - 1$, a design $d(N_0 + i + 1)$ with $N_0 + i + 1$ slides from a design $d(N_0 + i)$ with $N_0 + i$ slides by adding a slide to the latter. For each $i$, the slide added to $d(N_0 + i)$ compares a pair of treatment combinations chosen so as to minimize, over all the $p = \binom{v}{2}$ possibilities for such a pair, the value of the design criterion $\psi$, or equivalently of $\mathrm{tr}\{(X^T X)^{-1} W\}$ (see (2.10)), for the resulting design with $N_0 + i + 1$ slides. After $N - N_0$ steps, corresponding to $i = 0, 1, ..., N - N_0 - 1$, this yields and exact design $d(N)$ with $N$ slides.

The purpose of the *step-down* procedure, on the other hand, is to obtain an exact design with $N$ slides from an exact design, say $d(N_1)$, which involves $N_1$ ($> N$) slides and has a nonsingular in-formation matrix. It finds, for $i = 0, 1, ..., N_1 - N - 1$, a design $d(N_1 - i - 1)$ with $N_1 - i - 1$ slides from a design $d(N_1 - i)$ with $N_1 - i$ slides by deleting a slide from the latter. For each $i$, the slide deleted from $d(N_1 - i)$ is chosen, from amongst the $N_1 - i$ possibilities for such a slide, so as to minimize the value of the design criterion $\psi$, or equivalently of $\text{tr}\{(X^T X)^{-1} W\}$ (see (2.10)), for the resulting design with $N_1 - i - 1$ slides. After $N_1 - N$ steps, corresponding to $i = 0, 1, ..., N_1 - N - 1$, this yields and exact design $d(N)$ with $N$ slides

Thus, as hinted earlier, the addition or deletion in each step of step-up or step-down is done in the most efficient manner under the present design criterion. As an alternative to adding or deleting one slide in each step, one may wonder about the possibility of doing the entire augmentation or deletion in one shot. However, this often increases the computational burden very significantly. As an illustration, with reference to the $2^4$ factorial considered in the last subsection, consider the task of finding a 28-slide design starting from $d(48)$. Then step-down involves only 770 (= 48+47 +...+29) enumerations, whereas one shot deletion of 20 (= 48 − 28) slides in the most efficient manner requires as many as $\binom{48}{20}$, i.e., over $10^{13}$ enumerations. Because of this reason and also the fact that the present step-up/down procedure is itself capable of producing highly efficient designs, we do not any more consider the possibility of such one shot augmentation or deletion.

We are now in a position to formally describe our method of construction which combines approximate theory with step-up/down and consists of the following steps:

I. For a given parametrization and a given system of weights $w = (w_1, ..., w_n)$, apply the multiplicative algorithm described in (3.2) and (3.3) to obtain the optimal design measure $\tilde{\pi} = (\tilde{\pi}_1, ..., \tilde{\pi}_p)$.

II. Find the set $G$ of positive integers $g$ such that multiplication of $\tilde{\pi}_1, ..., \tilde{\pi}_p$ by a suitable constant and then rounding off to nearest integers yields an exact design which involves $g$ slides and has a nonsingular information matrix. Let $G = \{g_1, g_2, g_3, ...\}$, where $g_1, g_2, g_3, ...$ are in the increasing order. Let $d(g_1), d(g_2), d(g_3), ...$ be the corresponding exact designs as given by rounding off.

III. Suppose an exact design with a pre-assigned number, $N$, of slides is required. For $j = 1, 2, ...$, apply the step-up/down procedure to $d(g_j)$ so as to obtain an exact design, say $d_{0j}$, with $N$ slides. Step-up if $g_j < N$ and step-down if $g_j > N$. If $g_j = N$, then simply take $d_{0j} = d(g_j)$. Record the efficiency of $d_{0j}$. Go on increasing $j$ till no further gain in efficiency is observed. Take the $d_{0j}$ with the highest efficiency as the final exact design with $N$ slides.

An issue regarding Step III is whether one has to continue over a long range of $j$ so as to arrive at the $d_{0j}$ with the highest efficiency. Reassuringly, our computational experience shows that Step

III terminates quickly – typically upon or before reaching a $j$ such that $g_j$ approximately equals $2N$. This is natural because if $g_j$ is too large compared to $N$, then the starting design $d(g_j)$ is far apart from the target, which is why no further gain in efficiency is anticipated.

## 4. Examples

The examples in this section illustrate how our method of construction, summarized in I-III above, works under different parametrizations. In all these examples, even if rounding off to nearest integers yields an exact design with the stipulated $N$, i.e., $N$ belongs to the set $G$ introduced in II above, step-up/down gives a more efficient design. To save space, throughout we report only the best way of stepping up/down that yields an $N$-slide design with the highest efficiency; cf. III.

The exact designs obtained in the examples turn out to be highly efficient with efficiencies over 0.90 or even over 0.95. These efficiency figures, computed via (3.4), are relative to the optimal design measure $\tilde{\pi}$ which is not attainable in the exact setup. Thus they actually stand for lower bounds to the true efficiencies as exact designs. As a result, it is quite possible that some of the highly efficient designs obtained in this section, or indicated in the previous section, are actually optimal among all exact designs with the same number of slides.

*4.1 Examples under the baseline parametrization*

Examples 2-6 below refer to the baseline parametrization.

**Example 2**. Consider a $3^2$ factorial. Let $w = (1, 1)$. Then the optimal design measure $\tilde{\pi}$, arising from (3.2) and (3.3) is as shown in Table 1. Hence one can check that $G = \{12, 16, 18, 22, 30, …\}$

Table 1. *Optimal design measure in Example 2 under the baseline parametrization*

| Category | Pair | Mass |
|---|---|---|
| (i) | (01, 00), (02, 00), (10, 00), (20, 00) | 0.1054 |
| (ii) | (11, 01), (21, 01), (12, 02), (22, 02), (11, 10), (12, 10), (21, 20), (22, 20) | 0.0607 |
| (iii) | (02, 01), (20, 10) | 0.0242 |
| (iv) | (12, 11), (21, 11), (22, 12), (22, 21) | 0.0111 |
| (v) | Each remaining pair | 0 |

Let $N = 14$. We apply step-down from $d(16)$. Multiplication of the masses shown in Table 1 by 16, followed by rounding off to nearest integers, shows that $d(16)$ consists of the four pairs in category (i), each applied to two slides, and the eight pairs in category (ii), each applied to one slide. Step-down from $d(16)$ yields a 14-slide design with efficiency 0.9591 and consisting of the slides

(01, 00), (02, 00), (10, 00), (10, 00), (20, 00), (20, 00), (11, 01),
(21, 01), (12, 02), (22, 02), (11, 10), (12, 10), (21, 20), (22, 20).   □

**Example 3**. Consider a $3 \times 4$ factorial. Let $w = (1, 2)$. As in the last example, we can find the optimal design measure and check that $G = \{11, 17, 19, 22, 28, …\}$. Let $N = 18$. Step-down from $d(19)$ yields an 18-slide design with efficiency 0.9724 and consisting of the slides

(10, 00), (20, 00), (20, 00), (01, 00), (02, 00), (03, 00), (11, 01), (21, 01), (12, 02),
(22, 02), (13, 03), (23, 03), (11, 10), (12, 10), (13, 10), (21, 20), (22, 20), (23, 20).   □

**Example 4**. Consider a $2 \times 3^2$ factorial. Let $w = (1, 2, 2)$. Upon finding the optimal design measure, it is seen that $G = \{25, 26, 34, 38, 42,...\}$. Let $N = 29$. Step-down from $d(34)$ yields a 29-slide design with efficiency 0.9366 and consisting of the slides

(100, 000), (100, 000), (001, 000), (002, 000), (010, 000), (020, 000), (011, 001), (021, 001), (101, 001), (012, 002), (022, 002), (102, 002), (011, 010), (012, 010), (110, 010), (111, 011), (021, 020), (022, 020), (120, 020), (121, 021), (122, 022), (101, 100), (102, 100), (110, 100), (111, 101), (112, 102), (112, 110), (121, 120), (122, 120). . □

**Example 5**. Consider a $2^2 \times 4$ factorial. Let $w = (1, 1, 1)$. Upon finding the optimal design measure, it is seen that $G = \{25, 27, 30, 35, 37, ...\}$. Let $N = 30$. Step-down from $d(35)$ yields a 30-slide design with efficiency 0.9624 and consisting of the slides

(001, 000), (001, 000), (002, 000), (002, 000), (003, 000), (003, 000), (010, 000), (100, 000), (110, 010), (110, 010), (110, 100), (110, 100), (011, 001), (101, 001), (012, 002), (102, 002), (013, 003), (103, 003), (011, 010), (012, 010), (013, 010), (111, 011), (112, 012), (113, 013), (101, 100), (102, 100), (103, 100), (111, 101), (112, 102), (113, 103). □

**Example 6**. In all the preceding examples, the weights on interactions were equal to or greater than those on the main effects. Consider now a $2^4$ factorial, and for a change, consider the more traditional pattern of a decreasing sequence of weights as given by $w = (1, 1/2, 1/3, 1/4)$. Upon finding the optimal design measure, it is seen that $G = \{52, 56, 60, 72, ...\}$. Let $N = 27$. Step-down from $d(52)$ yields a 27-slide design with efficiency 0.9160 and consisting of the slides

(0001, 0000), (0001, 0000), (0010, 0000), (0010, 0000), (0100, 0000), (1000, 0000), (1000, 0000), (0011, 0001), (0101, 0001), (1001, 0001), (0011, 0010), (0110, 0010), (1010, 0010), (1011, 0011), (0101, 0100), (0110, 0100), (1100, 0100), (0111, 0101), (0111, 0110), (1001, 1000), (1010, 1000), (1100, 1000), (1101, 1001), (1110, 1010), (1111, 1011), (1101, 1100), (1111, 1110). □

*4.2 Examples under the all-to-next and hybrid parametrizations*

In Example 6, all factors have two levels and hence the baseline and all-to-next parametrizations are identical. So we revisit Examples 2-5 under the latter parametrization. Then the gaps caused by rounding off to nearest integers (see (a) of Subsection 3.2) are not very conspicuous. Still, however, step-up/down continues to remain very useful in producing designs with enhanced efficiency.

**Example 2** (continued). With a $3^2$ factorial, let $w = (1, 1)$ and $N = 14$. The optimal design measure $\tilde{\pi}$, arising from (3.2) and (3.3), is now as shown in Table 2.

Table 2. *Optimal design measure in Example 2 under the all-to-next parametrization*

| Category | Pair | Mass | Category | Pair | Mass |
|---|---|---|---|---|---|
| (i) | (02, 01), (20, 10) | 0.1118 | (vi) | (12, 02), (21, 20) | 0.0489 |
| (ii) | (01, 00), (10, 00) | 0.0991 | (vii) | (02, 00), (20, 00) | 0.0249 |
| (iii) | (12, 11), (21, 11) | 0.0789 | (viii) | (21, 01), (12, 10) | 0.0065 |
| (iv) | (11, 01), (11, 10) | 0.0632 | (ix) | (22, 02), (22, 20) | 0.0063 |
| (v) | (22, 12), (22, 21) | 0.0604 | (x) | Each remaining pair | 0 |

Hence one can check that $G = \{10, 12, 14, 16, 18, \ldots\}$. We apply step-up from $d(12)$. Multiplication of the masses shown above by 12, followed by rounding off to nearest integers, shows that $d(12)$ consists of the twelve pairs in (i)-(vi) each applied to one slide. Step-up from $d(12)$ yields a 14-slide design with efficiency 0.9481 and consisting of the slides

$$(20, 10), (20, 10), (01, 00), (02, 00), (10, 00), (02, 01), (11, 01),$$
$$(12, 02), (11, 10), (12, 11), (21, 11), (22, 12), (21, 20), (22, 21). \qquad \square$$

**Example 3** (continued). With a $3 \times 4$ factorial, let $w = (1, 2)$ and $N = 18$. Upon finding the optimal design measure, it is seen that $G = \{11, 12, 13, 14, 15, 16, 17, 18, 19, 20, \ldots\}$. Step-up from $d(17)$ yields an 18-slide design with efficiency 0.9673 and consisting of the slides

$$(01, 00), (10, 00), (02, 01), (11, 01), (03, 02), (12, 02), (13, 03), (11, 10), (20, 10),$$
$$(12, 11), (21, 11), (13, 12), (22, 12), (23, 13), (21, 20), (22, 21), (23, 22), (20, 00). \qquad \square$$

**Example 4** (continued). With a $2 \times 3^2$ factorial, let $w = (1, 2, 2)$ and $N = 29$. Upon finding the optimal design measure, it is seen that $G = \{28, 29, 30, 32, 34, \ldots\}$. Step-up from $d(28)$ yields a 29-slide design with efficiency 0.9467 and consisting of the slides

$$(001, 000), (010, 000), (100, 000), (002, 001), (011, 001), (101, 001), (012, 002), (102, 002),$$
$$(011, 010), (020, 010), (110, 010), (012, 011), (021, 011), (111, 011), (022, 012), (112, 012),$$
$$(021, 020), (120, 020), (022, 021), (121, 021), (101, 100), (110, 100), (102, 101), (120, 110),$$
$$(112, 111), (121, 111), (122, 112), (122, 121), (111, 101). \qquad \square$$

**Example 5** (continued). With a $2^2 \times 4$ factorial, let $w = (1, 1, 1)$ and $N = 30$. Upon finding the optimal design measure, it is seen that $G = \{24, 26, 28, 29, 30, 31, 32, 33, \ldots\}$. Step-down from $d(32)$ yields a 30-slide design with efficiency 0.9634 and consisting of the slides

$$(010, 000), (010, 000), (100, 000), (100, 000), (002, 001), (002, 001), (003, 002), (003, 002),$$
$$(001, 000), (011, 001), (101, 001), (012, 002), (102, 002), (013, 003), (103, 003), (011, 010),$$
$$(110, 010), (012, 011), (111, 011), (013, 012), (112, 012), (113, 013), (101, 100), (110, 100),$$
$$(102, 101), (111, 101), (103, 102), (112, 102), (113, 103), (111, 110). \qquad \square$$

Before concluding this section, we present an example which relates to the hybrid (2.5) of the baseline and all-to-next parametrizations. For this purpose, we visit Example 3 again.

**Example 3** (continued). With a $3 \times 4$ factorial, let $w = (1, 2)$ and $N = 18$. Consider the baseline parametrization for the three-level factor and the all-to-next parametrization for the four-level factor. Upon finding the optimal design measure, it is seen that $G = \{11, 13, 15, 17, 18, 20, 21, \ldots\}$. Step-up from $d(17)$ yields a 18-slide design with efficiency 0.9686 and consisting of the slides

$$(01, 00), (01, 00), (10, 00), (20, 00), (02, 01), (11, 01), (21, 01), (03, 02), (12, 02),$$
$$(22, 02), (13, 03), (23, 03), (11, 10), (12, 11), (13, 12), (21, 20), (22, 21), (23, 22). \qquad \square$$

## 5. Dye color effects

This section examines the consequence of including dye-color effects in the linear model (2.7). Let $\delta_R$ and $\delta_G$ denote the effects of red and green dye colors respectively. Then, for a slide which compares a pair of distinct treatment combinations $(j_1 \ldots j_n, j'_1 \ldots j'_n)$, the expected log intensity ratio equals $\{\delta_R + \tau(j_1...j_n)\} - \{\delta_G + \tau(j'_1...j'_n)\}$ if $j_1...j_n$ is colored red and $j'_1...j'_n$ is colored green, or $\{\delta_G + \tau(j_1...j_n)\} - \{\delta_R + \tau(j'_1...j'_n)\}$ if it is the other way around. Hence in the presence of dye-color effects, writing $\delta = \delta_R - \delta_G$, the expression for $E(Y)$ in (2.7) gets modified to

$$E(Y) = \delta q + X\theta, \qquad (5.1)$$

where $X$ and $\theta$ are the same as before and $q = (q_1,...,q_N)^T$, with $q_i = 1$ or $-1$ according as whether for the pair compared on the $i$th slide $(1 \leq i \leq N)$ the first member is colored red and the second colored green, or the first member is colored green and the second colored red.

We now proceed to define design efficiency in the presence of dye-color effects. Since $q^T q = N$, the information matrix for $\theta$ under (5.1) is given by

$$A = X^T (I_N - N^{-1} q q^T) X. \qquad (5.2)$$

Hence, analogously to (2.10), our design criterion $\psi$ equals $\sigma^2 \text{tr}(A^{-1}W)$. Now, as in Section 3, $X^T X = NM(\pi^{\text{exact}})$, where $\pi^{\text{exact}}$ is the design measure corresponding to the exact $N$-slide design under consideration. So by (5.2), $NM(\pi^{\text{exact}}) - A$ is nonnegative definite. Therefore,

$$\text{tr}(A^{-1}W) \geq N^{-1}\text{tr}[\{M(\pi^{\text{exact}})\}^{-1}W] \geq N^{-1}\text{tr}[\{M(\tilde{\pi})\}^{-1}W],$$

as before $\tilde{\pi}$ being the optimal design measure when dye-color effects are ignored. Hence in the spirit of (3.4), the efficiency of an exact $N$-slide design can now be defined as

$$\text{Eff(dye)} = \frac{N^{-1}\text{tr}[\{M(\tilde{\pi})\}^{-1}W]}{\text{tr}(A^{-1}W)}. \qquad (5.3)$$

Observe that (5.3) is even more conservative than (3.4) because it is relative to $\tilde{\pi}$ which is not only unattainable in the exact setup but also based on a more favorable model that ignores dye-color effects. Thus (5.3) actually represents a lower bound to the true efficiency as an exact design in the presence of dye-color effects. However, it provides a useful benchmark because if an exact design has a high value of Eff(dye) under a suitable dye-color assignment, then its true efficiency is even higher and it may as well be optimal among all exact designs with the same number of slides when dye-color effects included in the model. These points will be helpful when the examples of Section 4 are revisited later in this section.

Even designs, where every treatment combination is replicated an even number of times, have received attention in the context of dye-color effects. It is well-known (Kerr and Churchill, 2001) that an even design allows a dye-color assignment which is symmetric in the sense that each treatment combination is colored red and green equally often. It is not hard to see that then $X^T q = 0$, so that $A = X^T X$ by (5.2), and as a result, dye-color effects entail no loss of efficiency.

Often, however, the efficient designs obtained here, via approximate theory coupled with step-up/down, are not even. For instance, none of the designs in the examples of the last section is even. With a view to ensuring the robustness of such designs to dye-color effects, we now consider the idea of a nearly symmetric dye-color assignment which is applicable to any design. A dye-color assignment is called *nearly symmetric* if, for each treatment combination, the numbers of times it is colored red and green differ by at most unity. In the special case of a single factor, Schiffl (2011) found this kind of assignment to be very effective. In a multifactor situation as well, it is intuitively clear that under a nearly symmetric dye-color assignment, there will be very little loss of efficiency due to dye-color effects. Our findings, summarized in Table 3, amply testify to this point and thus reinforce what Schiffl (2011) observed in the single factor case. Satisfyingly, as shown below, a nearly symmetric assignment is possible for every design.

**Proposition 1**. *Every design allows a nearly symmetric dye-color assignment.*

*Proof.* Given any design $d$, let $J$ be the set of treatment combinations which are replicated an odd number of times in $d$. If $J$ is empty then $d$ is an even design. So it allows a dye-color assignment which is symmetric and hence nearly symmetric. Next, let $J$ be nonempty. Then the cardinality of $J$ is even as $d$ is a paired comparison design. Thus the treatment combinations in $J$ can be grouped into mutually exclusive and exhaustive pairs. We can interpret each such pair as a slide and then consider a design, say $d^*$, obtained by augmenting these slides to $d$. Evidently, then $d^*$ is an even design and hence it allows a symmetric dye-color assignment. It is easy to see that the induced dye-color assignment to the original design $d$, which is a subdesign of $d^*$, is nearly symmetric. □

We now revisit the examples of Section 4. For each design obtained in these examples, Table 3 displays a nearly symmetric dye-color assignment, by showing the dye-colors against each slide – e.g., (R01, G00) stands for a slide which compares the pair (01, 00) with 01 colored red and 00 colored green. The value of Eff(dye), as given by (5.2) and corresponding to the displayed dye-color assignment, is shown for each design. For ease in comparison, the value of Eff, computed via (3.4) and arising in the absence of dye-color effects, is also indicated. Throughout, Eff(dye) turns out to be quite close to the corresponding Eff, thus demonstrating that the loss of efficiency due to dye-color effects is very little under nearly symmetric dye-color assignment. Interestingly, this happens even for designs where a majority of treatment combinations are replicated an odd number of times. For instance, we have Eff(dye) = 0.9554 and Eff = 0.9673 for the design in Example 3 under the all-to-next parametrization, although as many as eight of the twelve treatment combinations have odd replication numbers. Even in absolute terms, the Eff(dye) values reported in Table 3 are high, all of them being greater than 0.90 or even 0.95. This is all the more impressive in view of the fact that Eff(dye) is only a conservative measure of the true efficiency, as discussed earlier.

Table 3. *Efficiencies of the designs in Examples 2-6 under nearly symmetric dye-color assignment*

| Example and parametrization | Design with nearly symmetric dye-color assignment | Eff(dye) | Eff |
|---|---|---|---|
| Example 2 Baseline | (R01, G00), (G02, R00), (R10, G00), (G10, R00), (R20, G00), (G20, R00), (R11, G01), (G21, R01), (G12, R02), (R22, G02), (G11, R10), (R12, G10), (R21, G20), (G22, R20). | 0.9481 | 0.9591 |
| Example 3 Baseline | (G10, R00), (R20, G00), (G20, R00), (R01, G00), (G02, R00), (R03, G00), (R11, G01), (G21, R01), (G12, R02), (R22, G02), (R13, G03), (G23, R03), (G11, R10), (R12, G10), (G13, R10), (R21, G20), (G22, R20), (R23, G20). | 0.9649 | 0.9724 |
| Example 4 Baseline | (R100, G000), (G100, R000), (R001, G000), (G002, R000), (R010, G000), (G020, R000), (R011, G001), (G021, R001), (R101, G001), (G012, R002), (R022, G002), (G102, R002), (G011, R010), (R012, G010), (R110, G010), (R111, G011), (R021, G020), (G022, R020), (G120, R020), (R121, G021), (G122, R022), (G101, R100), (R102, G100), (G110, R100), (G111, R101), (R112, G102), (G112, R110), (G121, R120), (R122, G120). | 0.9311 | 0.9366 |
| Example 5 Baseline | (R001, G000), (G001, R000), (R002, G000), (G002, R000), (R003, G000), (G003, R000), (R010, G000), (G100, R000), (R110, G010), (G110, R010), (R110, G100), (G110, R100), (R011, G001), (G101, R001), (R012, G002), (R102, G002), (R013, G003), (G103, R003), (R011, G010), (G012, R010), (R013, G010), (R111, G011), (G112, R012), (R113, G013), (G101, R100), (R102, G100), (G103, R100), (G111, R101), (R112, G102), (G113, R103). | 0.9602 | 0.9624 |
| Example 6 Baseline/ All-to-next | (R0001, G0000), (G0001, R0000), (R0010, G0000), (G0010, R0000), (R0100, G0000), (G1000, R0000), (R1000, G0000), (R0011, G0001), (G0101, R0001), (R1001, G0001), (G0011, R0010), (R0110, G0010), (R1010, G0010), (G1011, R0011), (R0101, G0100), (G0110, R0100), (R1100, G0100), (R0111, G0101), (G0111, R0110), (R1001, G1000), (R1010, G1000), (G1100, R1000), (R1101, G1001), (R1110, G1010), (G1111, R1011), (G1101, R1100). (R1111, G1110). | 0.9091 | 0.9160 |
| Example 2 All-to-next | (R20, G10), (G20, R10), (R01, G00), (G02, R00), (R10, G00), (R02, G01), (R11, G01), (G12, R02), (R11, G10), (R12, G11), (R21, G11), (G22, R12), (R21, G20), (R22, G21). | 0.9344 | 0.9481 |
| Example 3 All-to-next | (G01, R00), (G10, R00), (R02, G01), (R11, G01), (R03, G02), (R12, G02), (R13, G03), (R11, G10), (G20, R10), (G12, R11), (R21, G11), (R13, G12), (G22, R12), (R23, G13), (G21, R20), (R22, G21), (G23, R22), (R20, G00). | 0.9554 | 0.9673 |
| Example 4 All-to-next | (G001, R000), (R010, G000), (R100, G000), (R002, G001), (G011, R001), (G101, R001), (R012, G002), (R102, G002), (G011, R010), (R020, G010), (R110, G010), (G012, R011), (G021, R011), (R111, G011), (R022, G012), (G112, R012), (R021, G020), (R120, G020), (G022, R021), (R121, G021), (R101, G100), (G110, R100), (G102, R101), (G120, R110), (R112, G111), (G121, R111), (R122, G112), (G122, R121), (G111, R101). | 0.9431 | 0.9467 |

Table 3. (continued) *Efficiencies of the designs in Examples 2-6 under nearly symmetric dye-color assignment*

| Example and parametrization | Design with nearly symmetric dye-color assignment | Eff(dye) | Eff |
|---|---|---|---|
| Example 5 All-to-next | (R010, G000), (G010, R000), (R100, G000), (G100, R000), (R002, G001), (G002, R001), (R003, G002), (G003, R002), (R001, G000), (R011, G001), (G101, R001), (G012, R002), (R102, G002), (G013, R003), (R103, G003), (R011, G010), (G110, R010), (R012, G011), (R111, G011), (G013, R012), (R112, G012), (G113, R013), (R101, G100), (G110, R100), (G102, R101), (R111, G101), (R103, G102), (G112, R102), (R113, G103), (G111, R110). | 0.9597 | 0.9634 |
| Example 3 Hybrid | (R01, G00), (G01, R00), (G10, R00), (R20, G00), (G02, R01), (R11, G01), (G21, R01), (G03, R02), (R12, G02), (G22, R02), (G13, R03), (R23, G03), (G11, R10), (R12, G11), (R13, G12), (R21, G20), (G22, R21), (G23, R22). | 0.9577 | 0.9686 |

## 6. Concluding remarks

Before concluding, we dwell on a few points which influence our assumption that the log intensity ratios arising from different slides have the same variance and are uncorrelated.

*6.1 Biological variability*

We first consider the issue of biological variability and discuss it along the lines of Banerjee and Mukerjee (2008), but in the perspective of general factorials. In cDNA microarray experiments, the measurement error is typically swamped in biological variability. Thus the variance of an observed log intensity ratio arising from a slide which compares the pair $(j_1...j_n, j'_1...j'_n)$ can be expressed as $\gamma^2(j_1...j_n) + \gamma^2(j'_1...j'_n) + \lambda^2$, where $\gamma^2(j_1...j_n)$ and $\gamma^2(j'_1...j'_n)$ represent the biological variability within the cell populations given by $j_1...j_n$ and $j'_1...j'_n$, and $\lambda^2$ represents the variability due to the measurement error. If the variance components $\gamma^2(j_1...j_n)$ are equal for all cell populations (see e.g., Kerr, 2003, and Altman and Hua, 2006) with common value $\gamma^2$, then the log intensity ratios arising from different slides have a common variance $\sigma^2 = 2\gamma^2 + \lambda^2$. In this case, if independent biological replications are used throughout the experiment then these ratios are also uncorrelated and all our results go through with $\sigma^2 = 2\gamma^2 + \lambda^2$.

The assumption of equal variance, however, breaks down if the $\gamma^2(j_1...j_n)$ are not all equal. In this situation, let $\breve{\gamma}^2(j_1...j_n) = \gamma^2(j_1...j_n)/\lambda^2$ for each $j_1...j_n$. Then, writing $p = \binom{v}{2}$ as before, for any $k$ $(1 \leq k \leq p)$, the variance of the log intensity ratio arising from a slide that compares the pair of treatment combinations labeled $k$ is proportional to $a_k^2$, where $a_k^2 = \breve{\gamma}^2(j_1...j_n) + \breve{\gamma}^2(j'_1...j'_n) + 1$, if this pair equals $(j_1...j_n, j'_1...j'_n)$. Thus, if only independent biological replications are used, then under this heteroscedastic but uncorrelated scenario, the optimal design measure, associated with the generalized least squares (GLS) estimator of the parametric vector $\theta$ representing the main effects and interactions, can again be obtained via the multiplicative algorithm, with $x_k$ and $M(\pi)$ in (3.2) and (3.3) being replaced respectively by $a_k^{-1} x_k$ and $M^{\#}(\pi) = \Sigma_{k=1}^{p} \pi_k a_k^{-2} x_k x_k^T$. If $\pi^{\#}$ is the optimal design measure so obtained, then as a counterpart of (3.4), the efficiency of an exact $N$-slide design $d$ can be assessed by

$$\text{Eff}^{\#} = \frac{N^{-1}\text{tr}[\{M^{\#}(\pi^{\#})\}^{-1} W]}{\text{tr}[\{(X^T X)^{-1} X^T V X (X^T X)^{-1}\} W]},$$

where $X$ is the design matrix of $d$ (cf. (2.7)) and $V$ is an $N \times N$ diagonal matrix with $i$th diagonal element ($1 \leq i \leq N$) equal to $a_k^2$ if the $i$th slide of $d$ compares the pair labeled $k$. Note that even though $\pi^{\#}$ relates to the GLS estimator of $\theta$, the denominator of $\text{Eff}^{\#}$ corresponds to the ordinary least squares (OLS) estimator of $\theta$ as obtained from $d$. This makes $\text{Eff}^{\#}$ is rather conservative but realistic because the ratios $\breve{\gamma}^2(j_1...j_n)$ are unknown in practice, and hence it makes sense to assess the performance of an exact design on the basis of only the OLS estimator arising from it.

In order to examine the robustness of our findings to heteroscedasticity as discussed above, we first revisit Example 3 which concerns a $3 \times 4$ factorial, with $w = (1, 2)$ and $N = 18$. Three patterns are considered for

$$(\breve{\gamma}^2(00), \breve{\gamma}^2(01), \breve{\gamma}^2(02), \breve{\gamma}^2(03), \breve{\gamma}^2(10), \breve{\gamma}^2(11),$$
$$\breve{\gamma}^2(12), \breve{\gamma}^2(13), \breve{\gamma}^2(20), \breve{\gamma}^2(21), \breve{\gamma}^2(22), \breve{\gamma}^2(23)),$$

namely, (i) (2, 2.5, 2.5, 3, 2.5, 3, 3, 4, 3, 3.5, 4, 4.5), (ii) (2, 3, 3, 4, 3, 4, 5, 6, 5, 6, 7, 8) and (iii) (8, 7, 6, 5, 6, 5, 4, 3, 4, 3, 3, 2). Under (i)-(iii), the design reported in Example 3 under the baseline parametrization is found to have $\text{Eff}^{\#}$ values 0.9734, 0.9647 and 0.9375 respectively, and the design reported in the same example under the all-to-next parametrization is seen to have $\text{Eff}^{\#}$ values 0.9692, 0.9625 and 0.9308 respectively. Turning next to Example 5 which concerns a $2^2 \times 4$ factorial, with $w = (1, 1, 1)$ and $N = 30$, we consider three patterns for

$$(\breve{\gamma}^2(000), \breve{\gamma}^2(001), \breve{\gamma}^2(002), \breve{\gamma}^2(003), \breve{\gamma}^2(010), \breve{\gamma}^2(011), \breve{\gamma}^2(012), \breve{\gamma}^2(013),$$
$$\breve{\gamma}^2(100), \breve{\gamma}^2(101), \breve{\gamma}^2(102), \breve{\gamma}^2(103), \breve{\gamma}^2(110), \breve{\gamma}^2(111), \breve{\gamma}^2(112), \breve{\gamma}^2(113)),$$

namely, (i) (2, 3, 4, 5, 3, 4, 4, 6, 4, 5, 6, 7, 5, 6, 7, 8), (ii) (3, 4, 5, 7, 4, 6, 7, 8, 5, 7, 8, 9, 6, 8, 10, 11) and (iii) (11, 10, 8, 6, 9, 8, 7, 5, 8, 7, 6, 4, 7, 5, 4, 3). Under (i)-(iii), the design reported in Example 5 under the baseline parametrization has $\text{Eff}^{\#}$ values 0.9544, 0.9536 and 0.9360 respectively, and the design reported in the same example under the all-to-next parametrization has $\text{Eff}^{\#}$ values 0.9483, 0.9449 and 0.9426 respectively. The picture is equally encouraging for the designs obtained in the other examples too. Thus the highly efficient designs obtained by our method turn out to be quite robust to heteroscedasticity as may be caused by biological variability.

*6.2 Technical replication*

The log intensity ratios arising from different slides become correlated when the same biological sample is applied to two or more slides, i.e., technical replication is permitted. The design problem in this situation becomes rather complex. As noted in Sanchez and Glonek (2009), this happens on three counts:

(i) The design space becomes much larger than here. In addition to deciding on the pairs to be compared on the available slides and their dye-coloring, one has to decide on technical versus biological replication for each treatment combination which is replicated more than once, and the number of possibilities regarding the latter grows fast as the number of replications increases.

(ii) Even if the log-intensity ratios remain homoscedastic, they now get correlated and the correlation is unknown, thus complicating matters.

(iii) Third, although the number of slides is typically the most important constraint in microarray experiments (see e.g., Bueno Filho et al., 2006), there may also be constraints on the numbers biological samples available for the different treatment combinations, i.e., cell populations, and so some amount of technical replication may be unavoidable.

If the point in (iii) is ignored, then the heuristic studies in Banerjee and Mukerjee (2008), which reinforce some earlier findings in Kerr (2003) and Kendziorski et al. (2005), suggest in favor of having only biological replications from the point of view of efficiency. However, the mathematics underlying a complete treatment of this issue appears to be daunting at this stage. As for (ii), along the lines of the last subsection, one may think of using approximate theory to conduct robustness studies. Finally, if the point in (iii) is of importance in a given context, then in the spirit of Kerr (2003), it makes sense to formulate the design problem in terms of a cost function which incorporates the cost of the slides as well as that of biological replication.

A more comprehensive attempt towards addressing the points in (i)-(iii) is beyond the scope of this article. We conclude with the hope that the present work will generate interest in these and related issues.

**Acknowledgement**. The work of RZ was supported by NNSF of China grants 11171165 and 10871104. The work of RM was supported by the J.C. Bose National Fellowship of the Government of India and also a grant from the Indian Institute of Management Calcutta.